\documentclass[a4paper,12pt]{article}
\pdfoutput=1

\usepackage{amsmath,amssymb}
\usepackage[bookmarks=true]{hyperref}
\usepackage[nosort]{cite}
\usepackage[bulletsep]{collect}
\def\gfxon{\usepackage[final]{graphicx}}

\gfxon

\usepackage{rotating}

\ifx\hypersetup\sadfkjashdfkxja\else
\hypersetup{pdftitle={Open Perturbatively Long-Range Integrable gl(N) Spin Chains}}
\hypersetup{pdfauthor={N. Beisert, F. Loebbert}}
\hypersetup{pdfsubject={Mathematical Physics}}
\hypersetup{pdfkeywords={Long-Range Spin Chain, Open Boundary Conditions}}
\hypersetup{plainpages=false}
\hypersetup{bookmarksnumbered=true}
\hypersetup{pdfstartview=FitH}
\hypersetup{pdfpagemode=None}
\hypersetup{colorlinks=false}
\hypersetup{citebordercolor={.5 1 .5}}
\hypersetup{urlbordercolor={.5 1 1}}
\hypersetup{linkbordercolor={1 .7 .7}}
\fi

\numberwithin{equation}{section}

\makeatletter
\let\old@makecaption=\@makecaption
\def\@makecaption{\small\old@makecaption}
\makeatother

\def\[{\begin{equation}}
\def\]{\end{equation}}
\def\<{\begin{eqnarray}}
\def\>{\end{eqnarray}}

\newcommand{\nn}{\nonumber}
\newcommand{\nln}{\nonumber\\}
\newcommand{\nl}[1][0pt]{\nonumber\\[#1]&\hspace{-4\arraycolsep}&\mathord{}}

\newcommand{\earel}[1]{\mathrel{}&\hspace{-2\arraycolsep}#1\hspace{-2\arraycolsep}&\mathrel{}}
\newcommand{\eq}{\earel{=}}

\ifx\genfrac\sdflkaj
\newcommand{\atopfrac}[2]{{{#1}\above0pt{#2}}}
\else
\newcommand{\atopfrac}[2]{\genfrac{}{}{0pt}{}{#1}{#2}}
\fi
\newcommand{\sfrac}[2]{{\textstyle\frac{#1}{#2}}}
\newcommand{\half}{\sfrac{1}{2}}
\newcommand{\ihalf}{\sfrac{i}{2}}

\newcommand{\et}[1]{\mathrm{#1}}

\newcommand{\ham}{\mathcal{H}}
\newcommand{\charge}{\mathcal{Q}}

\newcommand{\indups}[1]{_{\mathrm{\scriptscriptstyle #1}}}

\newcommand{\rep}[1]{{\mathbf{#1}}}

\newcommand{\alg}[1]{\mathfrak{#1}}

\newcommand{\lrbrk}[1]{\left(#1\right)}
\newcommand{\bigbrk}[1]{\bigl(#1\bigr)}

\newcommand{\comm}[2]{[#1,#2]}

\newcommand{\ket}[1]{|#1\rangle}

\newcommand{\Perm}[1]{[#1]}

\newdimen\yysquaresize
\newdimen\yyrsquaresize
\newdimen\yythickness
\newdimen\yyskip
\def\yysquare#1{%
\setlength{\yyrsquaresize}{\yysquaresize}%
\addtolength{\yyrsquaresize}{-2\yythickness}%
\vrule width \yythickness%
\vbox to \yysquaresize{%
  \hrule height \yythickness\vss%
  \hbox to \yyrsquaresize{\hss#1\hss}%
  \vss\hrule height \yythickness}%
\vrule width \yythickness}

\def\yyyoung#1{\vtop{\baselineskip0pt\lineskip-\yythickness\halign{\tabskip-\yythickness&\yysquare{##}\cr #1}}}

\newcommand{\young}[1]{\hskip\yyskip\mbox{\yyyoung{#1}}\hskip\yyskip}

\ifx\href\asklfhas\newcommand{\href}[2]{#2}\fi
\newcommand{\arxivlink}[1]{\href{http://arxiv.org/abs/#1}{arxiv:#1}}

\begin{document}
\pagenumbering{roman}

\thispagestyle{empty}
\begin{flushright}\footnotesize
\texttt{\arxivlink{0805.3260}}\\
\texttt{AEI-2008-032}%
\end{flushright}
\vspace{0.5cm}

\begin{center}
{\Large\textbf{\mathversion{bold}%
Open Perturbatively Long-Range\\Integrable
$\alg{gl}(N)$ Spin Chains}\par}
\vspace{1cm}

\textsc{N.~Beisert and F.~Loebbert}
\vspace{5mm}

\textit{Max-Planck-Institut f\"ur Gravitationsphysik\\
Albert-Einstein-Institut\\
Am M\"uhlenberg 1, D-14476 Potsdam, Germany}\vspace{3mm}

\texttt{nbeisert@aei.mpg.de}\\
\texttt{florian.loebbert@aei.mpg.de}\\
\par\vspace{1cm}

\vfill

\textbf{Abstract}\vspace{5mm}

\begin{minipage}{12.7cm}
We construct the most general perturbatively long-range
integrable spin chain with spins transforming
in the fundamental representation of $\alg{gl}(N)$
and \emph{open boundary conditions}.
In addition to the previously determined bulk moduli
we find a new set of parameters determining
the \emph{reflection phase shift}.
We also consider \emph{finite-size} contributions
and comment on their determination.
\end{minipage}

\end{center}
\vfill

\newpage
\setcounter{page}{1}
\pagenumbering{arabic}
\renewcommand{\thefootnote}{\arabic{footnote}}
\setcounter{footnote}{0}

\tableofcontents
\vspace{5mm}

\section{Introduction}

During the past few years a long list of evidence
has been collected that the spectrum
of planar anomalous dimensions
in $\mathcal{N}=4$ supersymmetric gauge theory
is equivalent to the energy spectrum
of a certain integrable spin chain
\cite{Minahan:2002ve,Beisert:2003tq,Beisert:2003yb}.
The spectrum of this spin chain is described efficiently
by Bethe equations, see \cite{Beisert:2005fw}
and references therein,
at least for sufficiently long chains.
This spin chain model is special in many respects.
Most importantly the interactions are genuinely non-local
and between multiple sites of the chain \cite{Beisert:2003tq}.

Our knowledge of such integrable long-range chains is still very limited.
The best known exception is the Inozemtsev
chain which is similar to the Haldane--Shastry chain but
with elliptic rather than trigonometric
dependence on the separation of spins.
The situation for the Inozemtsev chain is similar to
the $\mathcal{N}=4$ gauge theory chain:
The asymptotic spectrum (for long chains) is described by Bethe equations
while determination of equations for the exact (finite-length) spectrum
remains a challenge.

Although the $\mathcal{N}=4$ gauge theory chain is similar to the
Inozemtsev chain \cite{Serban:2004jf},
it belongs to a more general class of long-range chains
with simultaneous interactions between more than two spins.
In the full $\mathcal{N}=4$ gauge theory chain the interactions
can even change the number of spin sites \cite{Beisert:2003ys},
but here we shall focus on the $\alg{su}(2)$ sector of the model where the
length-fluctuations are frozen out.
The setup for such \emph{perturbatively} long-range%
\footnote{The name \emph{long-range} spin chain actually refers to the fact
that for \emph{finite} coupling $\lambda$ the interactions are indeed of infinite,
i.e.\ long range. However, considering the problem within perturbation theory
all relevant contributions have finite, i.e.\ short range.
Even though it might therefore be more appropriate
to speak of \emph{short-range} chains
we stick to the notion of long-range chains for historical reasons.}
chains was outlined in \cite{Beisert:2003tq}.
In the simplest case, it is a chain with spins transforming in the
fundamental (spin-$\half$) representation of $\alg{su}(2)$.
More degrees of freedom corresponding to a deformation \cite{Arutyunov:2004vx}
of the overall magnon scattering phase
were later discovered in \cite{Beisert:2004jw}.
A full treatment of all deformation parameters
in the more general setting of $\alg{gl}(N)$ with fundamental spins
and up to interactions of range $6$ was conducted in \cite{Beisert:2005wv}.
It showed that, independently of the rank $N$,
there are four types of deformation parameters:
The parameters $\alpha_\ell(\lambda)$ govern
deformations of the dispersion relation.
The parameters $\beta_{r,s}(\lambda)$ correspond to deformations
of the overall magnon scattering phase.
The parameters $\gamma_{r,s}(\lambda)$ fix the linear combinations
of commuting charges
in terms of some canonical basis.
And finally the parameters $\epsilon_{\ell,n}(\lambda)$
correspond to similarity transformations of the integrable system
without impact on the spectrum.
Here the coupling constant $\lambda\approx 0$
controls the range of the Hamiltonian:
A contribution at order $\lambda^\ell$
is allowed to have interactions among $\ell+2$ neighboring sites.
Complete integrability of this system was initially only a
conjecture based on the existence of \emph{one} conserved charge.
Later it was shown that
the Hamiltonian possesses Yangian symmetry
which constitutes a formal proof of integrability \cite{Beisert:2007jv}.

In this paper we shall consider \emph{open} spin chains
which is a natural generalization of the above closed chains.
In the original gauge theory setup open boundaries correspond
most naturally to ``quarks'',
i.e.\ to fields transforming in the \emph{fundamental}
(as opposed to adjoint) representation of the gauge group
\cite{Chen:2004mu,DeWolfe:2004zt,Erler:2005nr}.
On the string theory side of the AdS/CFT correspondence, the quarks are
represented by strings ending on D-branes
\cite{Stefanski:2003qr,Susaki:2004tg,Susaki:2005qn,Okamura:2005cj,Mann:2006rh,Ahn:2007bq}.
However, even in a gauge theory with adjoint fields only, such as
$\mathcal{N}=4$ SYM, open spin chains make an appearance:
For example one can turn
gauge covariant local operators
representing open chains into a non-local gauge invariant object
by means of a Wilson loop \cite{Drukker:2006xg}.
Alternatively an open chain (open string)
can end on a determinant-like local operator (giant graviton)
\cite{Berenstein:2005vf,Berenstein:2006qk,Agarwal:2006gc,Hofman:2007xp,Agarwal:2007mq,Ahn:2008df}.

Here we would like to perform an exhaustive study of
\emph{open} perturbatively long-range
integrable spin chains with $\alg{gl}(N)$ symmetry
analogously to the one for closed chains in \cite{Beisert:2005wv}.
Our aim is to understand how the long-range
interactions can deform the boundary conditions as well as what the restrictions for the bulk Hamiltonian are.
For practical reasons we will not study the most general boundary conditions,
but only those which preserve manifest $\alg{gl}(N)$ invariance.
Note that in many physical models,
such as most of the systems discussed above,
the bulk symmetry is actually broken by the boundary conditions,
e.g.\ $\alg{gl}(N)\to\alg{gl}(M)\times\alg{gl}(N-M)$.
Therefore our results do not apply directly to these models,
but we expect that the qualitative picture will be
roughly the same as for our symmetry-preserving boundary conditions.
Indeed previous results on concrete open perturbatively long-range chain models
\cite{Arnaudon:2003zw,Arnaudon:2003gj,McLoughlin:2005gj,Okamura:2006zr,Chen:2007ec}
confirm this expectation.

Our procedure is the same as in \cite{Beisert:2005wv}:
We will make a general ansatz for
two long-range spin chain operators including open boundary terms.
By demanding that the two commute we obtain a tentatively
integrable system.
Unfortunately, Yangian symmetry is broken by the
boundary conditions in nearest-neighbor spin chains
even if they are integrable.
Therefore we cannot use Yangian symmetry to provide a
formal proof of integrability of our boundary terms.
We will then perform the coordinate Bethe ansatz and obtain
asymptotic Bethe equations for long-range open chains.
These describe the spectrum of the perturbatively long-range Hamiltonian
(if indeed the assumed integrability holds).

\section{The Spin Chain Model}

We consider an open spin chain with spins transforming
in the fundamental representation of $\alg{gl}(N)$.
A spin chain state of given length $L$ is an element
of the tensor product space $(\mathbb{C}^N)^{\otimes L}$.
A basis for such states is given by
\[
 \ket{a_1,\dots, a_L},\hspace{10mm} a_k=1,\dots,N,
\]
where each $a_k$ represents one basis vector of $\mathbb{C}^N$.
Each homogeneous $\alg{gl}(N)$ invariant local operator can
be built from permutations of adjacent sites.
We write these permutations in the form $\Perm{a_1,\dots,a_n}$
acting homogeneously on a spin chain of length $L$ by
\[\label{eq:terms}
 \Perm{a_1,\dots,a_n}\ket{b_1,\dots,b_L}=\sum_{k=1}^{L-n+1}\ket{b_1,\dots,b_k,b_{k+a_1},\dots,b_{k+a_n},b_{k+n+1},\dots,b_L}.
\]
For example each spin chain state is an eigenstate
of the operator $\Perm{1}$ with eigenvalue $L$.
Note that for a closed chain, the operator $\Perm{1,2}$
would have the same property.
For an open chain its eigenvalue is $L-1$
instead as will be explained below.

We can thus introduce the length operator $\mathcal{L}$
counting the number of spin sites in the chain.
Furthermore we define the operator $\mathcal{B}$
to measure whether the chain has a boundary or not,
i.e.\ to vanish on a space of closed spin chain states
and to equal $1$ on a space of open chains.
In our notation we have
\begin{align}
 \mathcal{L}&=\Perm{1},\nln
\mathcal{B}&=\Perm{1}-\Perm{1,2}.
\end{align}
These operators commute with all local operators.

\paragraph{Boundary Terms.}

\begin{figure}[t]\centering
\includegraphics{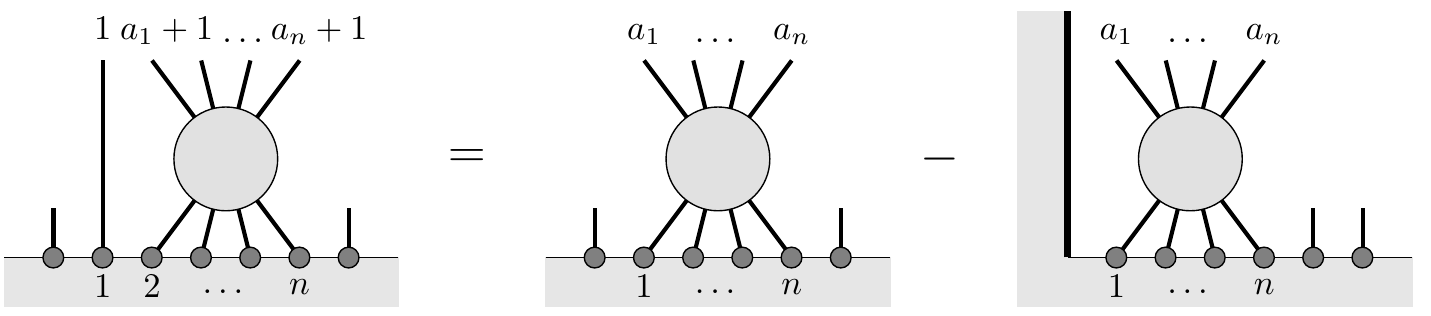}
\caption{The two interactions contributing to a
boundary term act differently only at the boundaries.}
\label{FigBoundTerms}
\end{figure}

The new feature of open spin chains as compared
to the closed chains are \emph{boundary terms}
enlarging the set of building blocks for the $\alg{gl}(N)$
invariant operators.
This stems from the fact that spectator legs may not be
dropped by identifying certain states since these states act
differently on the boundaries as illustrated
in Figure~\ref{FigBoundTerms}
\[
 \Perm{1,a_1+1,\dots,a_n+1}\neq\Perm{a_1,\dots,a_n}\neq\Perm{a_1,\dots,a_n,n+1}.
\]
Curiously these boundary terms can be encoded into homogeneous bulk terms,
i.e.\ structures of the form
\begin{align}
 &\Perm{1,a_1+1,\dots,a_n+1}-\Perm{a_1,\dots,a_n},\nln
 &\Perm{a_1,\dots,a_n,n+1}-\Perm{a_1,\dots,a_n},\qquad n>1,
\end{align}
only act on the boundaries and vanish identically in the bulk.

\paragraph{Finite Length Terms.}

When spectator legs occur on both sides of the operator
we find another interesting structure. Operators of the form
\begin{align}\label{flform}
&\Perm{1,a_1+1,\dots,a_n+1,n+2}-\Perm{1,a_1+1,\dots,a_n+1}\nln
&-\Perm{a_1,\dots,a_n,n+1}+\Perm{a_1,\dots,a_n},\hspace{35mm} n>2,
\end{align}
exclusively act on states of length $n$
and vanish identically on the rest of the space of states.
We are not going to include these terms into our considerations,
i.e.\ we replace terms of the form $\Perm{1,a_1+1,\dots,a_n+1,n+2}$
by terms of shorter range according to \eqref{flform}
at every point of our computation.

Note however that these structures provide a notation
for operators acting on chains of one specific length
only and are therefore useful for addressing the
problem of finite length effects in perturbative
long-range chains.
We analyze this issue in Section~\ref{finiteL}.

\begin{table}[t]\centering
\begin{tabular}{|l|c|c|c|c|}\hline
Type of structures & Bulk&Boundary&Finite Length&All\\ \hline
Number of structures &$R!-(R-1)!+1$&$2(R-1)!-1$&$\sum\limits_{l=3}^R (l-2)!$&$\sum\limits_{l=1}^R l!$\\ \hline
\end{tabular}
\caption{Numbers of the different kinds of building blocks
with interaction range $R$ in the ansatz for a generic local operator
in this problem.}
\end{table}

\paragraph{Parity.}

We introduce a parity operation acting on fundamental interactions as
\[
 P \,\Perm{a_1,\dots,a_n} \,P^{-1}=\Perm{n-a_n+1,n-a_{n-1}+1,\dots,n-a_1+1}.
\]
It is useful to classify the interactions according to their parity,
e.g.\ if one is interested in a parity conserving model.

\section{Constructing Conserved Charges}
\label{sec:constr}

The integrability of a spin chain model is expressed
by the existence of an infinite tower of conserved charges,
all commuting among each other. The idea of a \emph{perturbatively}
integrable spin chain is that these commuting charges
are expressible as a perturbation series in a small parameter $\lambda$
\cite{Beisert:2003tq}
\[
 \charge_{r}=\sum_{k=0}^{\infty} \lambda^k \charge_{r}^{(k)},\hspace{5mm}r=1,2,\dots.
\]
Here the $\charge_{2r}^{(k)}$ have maximal interaction range $2r+k$,
i.e.\ with each power of $\lambda$ the range of the charge increases by one.%
\footnote{We assume the length of the spin chain to exceed
the length of the considered operators.
Hence, our analysis will not necessarily apply
to short chains giving rise to spanning interactions,
see Section~\ref{finiteL}.}

It is well known that for open spin chains only half of the
integrable charges present for closed chains are conserved.
It turns out that the \emph{odd charges} $\charge_{2r+1}$
commute with the \emph{even charges} $\charge_{2s}$
only up to boundary terms.
Therefore the tower of commuting
charges can only be constructed of the even charges $\charge_{2r}$
with an even interaction range $2r$ at zeroth order in $\lambda$.

Making the most general ansatz for the charges with index $2r$,
i.e.\ a linear combination of all local operators
of length $\leq 2r+k$ with arbitrary coefficients, we require that
\[\label{commchar}
  \comm{\charge_{2r}}{\charge_{2s}}=\mathcal{O}(\lambda^{k+1}),\hspace{5mm} r,s=1,2,\dots
\]
and solve order by order for the coefficients in the ansatz.
In this paper we explicitly construct the first two commuting
charges $\charge_2$ and $\charge_4$, where $\charge_2$
is defined to be the Hamiltonian of the system.
Note that technically this is merely a necessary but not a sufficient
condition for integrability. For a closed chain,
however, the existence of one conserved charge
$\charge_3$ has experimentally turned out
to be a sufficient condition for integrability
\cite{Beisert:2005wv,Beisert:2007jv}.
The reason for this behavior is unclear, but we expect the
same to be true for open chains.

The construction is analogous to the one for closed chains
in \cite{Beisert:2005wv}. It is somewhat more challenging
because the range of $\charge_4$
is one step longer than the one of $\charge_3$
and consequently the ansatz contains many more structures at
a given perturbative order. We present the result
for the Hamiltonian $\charge_2$ up to second order
in Table~\ref{tab:Q2} at the end of the paper.
We have also constructed $\charge_2$ and $\charge_4$ up to third order,
but the resulting expressions are too lengthy
(and not enlightening) to be presented here.
Our result agrees with the one for closed chains \cite{Beisert:2005wv,Beisert:2007jv}
when projecting out the boundary contributions.
Note however that some of the closed chain parameters
are not present for open chains
because they are incompatible with the boundary conditions.

\section{Asymptotic Bethe Ansatz}

We now perform the asymptotic Bethe ansatz for the open $\alg{gl}(N)$ spin chain.
For closed chains the general ansatz was presented in \cite{Sutherland:1978aa,Staudacher:2004tk}.
There the idea was to construct eigenstates of the Hamiltonian for an infinitely
long chain as a superposition of asymptotic $n$-particle states.
The zero-particle states yield the vacuum energy while the one-particle states
determine the dispersion relation $E(p)$.
Two-particle states with momenta $p$ and $q$ then fix the
scattering matrices $S(p,q)$ in flavor space such that one ends
up with a two-particle eigenstate of the Hamiltonian.
Integrability implies that the scattering of several
particles with many different flavors reduces to the two situations
where particles of either different or the same flavor scatter with each other.
Considering a perturbatively long-range Hamiltonian $\mathcal{H}(\lambda)$
the particles can see each other already at finite distances.
This leads to a scattering matrix which depends on the distance
of the magnons in contrast to a sharp change of phase for the zero-order Hamiltonian.

Imposing closed periodicity conditions, i.e.\ the closed Bethe equations,
on the infinitely long chain then requires that shifting a particle by $L$
sites yields a phase factor $e^{ipL}$ which is to be
equal to the product of scattering matrices corresponding to the interactions
with the other excitations on the chain.

\begin{figure}[t]\centering
\includegraphics{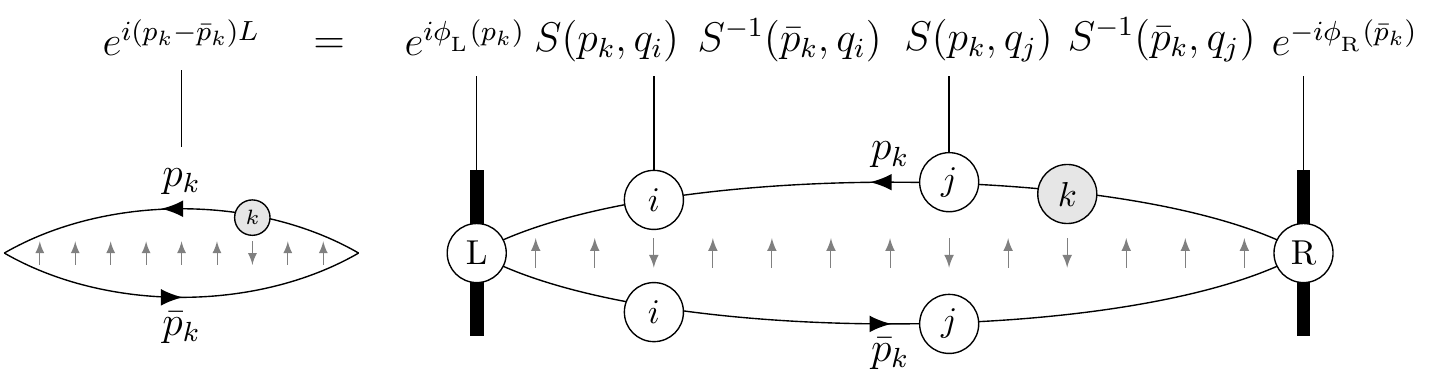}
\caption{Boundary Bethe equations:
A particle of momentum $p_k$ scatters with an excitation
of momentum $q_j$ and $q_i$.
The particle is reflected at the left boundary
where its momentum changes to $\bar p_k$ and
it picks up a boundary scattering phase.
Having scattered with the particles $i$ and $j$ again,
this time with a different momentum $\bar p_k$,
the particle acquires another phase factor at the right boundary
and the momentum changes back $\bar p_k\to p_k$.
After this period of $2L$ sites, associated with a phase factor
of $e^{i(p_k-\bar p_k)L}$, the magnon has returned
to its original position, moving in its original direction.}
\label{FigBoundaryBethe}
\end{figure}

For the open spin chain these periodicity conditions change:
In order for the particle to arrive at the same position on the chain,
moving in the same direction, it has to be shifted by $2L$ sites.
On this way the magnon is reflected at the two boundaries
and passes all other excitations on the chain twice.
Each boundary gives rise to a \emph{boundary scattering phase}
such that the difference of these phases is part of the Bethe equations.
Furthermore the momentum $p$ of the ingoing particle changes to $\bar p$
after reflection.%
\footnote{Note that we have $\bar p\neq -p$ since the considered
Hamiltonian does in general not preserve parity.}
The two momenta are the two solutions of the equation $E(p)=E(\bar p)$
related by the \emph{reflection map}.
For periodicity we therefore have to require
that the phase factor $e^{i(p-\bar p) L}$
equals the product of scattering matrices $S(p,q_j)$
and $S(\bar p,q_j)$ with the boundary phase factor
$e^{2i\phi(p)}$ as illustrated in Figure~\ref{FigBoundaryBethe}.

Again, integrability implies that the multi-particle problem reduces
to two-particle scattering and single particle interactions with the boundaries.
Therefore it suffices to consider a one-particle state,
with a single flavor due to $\alg{gl}(N)$ invariance,
in order to determine the boundary scattering phase and the reflection map $\bar p(p)$.
The bulk scattering matrices $S(p,q)$ can be computed in the same fashion
as for the closed spin chain.

To determine the boundary quantities we require the state
\[
	\ket{\Phi\indups{L/R}}=\sum_{x=1}^L \Phi\indups{L/R}(x) \ket{x},
\]
with
\begin{align}\label{boundaryansatz}
	\Phi\indups{L}(x)&=e^{i p x}+  e^{2i\phi\indups{L}(p)} e^{i \bar p (x-1)}+a\indups{L}(x),\nln
	\Phi\indups{R}(x)&=e^{i p (x-L)}+ e^{2i\phi\indups{R}(p)} e^{i \bar p (x-1-L)}+a\indups{R}(x),
\end{align}
to be an eigenstate of the Hamiltonian close to either the left or the right boundary.
That is to say that we project the eigenvalue equation
\[
\ham \ket{\Phi\indups{L/R}}=E(p) \ket{\Phi\indups{L/R}}
\]
on all states within the interaction range of the boundary at current perturbative order
in $\lambda$ and solve the resulting system of equations.
This defines the boundary scattering phases $\phi\indups{L/R}(p)$
as well as the reflection map $\bar p(p)$
as a perturbation series in $\lambda$.
Furthermore we have to solve for
the local boundary parameters $a\indups{L}(x)$ and $a\indups{R}(x)$ which,
however, represent UV physics and will not be of further interest in this paper
\[
a\indups{L}(x)=\mathcal{O}(\lambda^x),\hspace{10mm} a\indups{R}(x)=\mathcal{O}(\lambda^{L-x+1}).
\]

After determining all physically relevant scattering factors for the problem,
we can impose the periodicity conditions for the considered spin chain type.
We first present the Bethe equations for the open $\alg{gl}(N)$ spin chain
and afterwards discuss the different kinds of free parameters of the system.
Each state is described by a set of Bethe roots $u_{\ell,k}$.
The label $\ell=1,\dots,N-1$ indicates the flavor of the Bethe root
whereas the label $k=1,\dots,K_{\ell}$ indexes the set of Bethe roots of flavor $\ell$.
The main Bethe equation at level $\ell=1$ reads
\begin{align}\label{BetheEquations}
1=&\lrbrk{\frac{x_{1,k}^-}{x_{1,k}^+}}^L
    	 \prod_{\textstyle\atopfrac{j=1}{j\neq k}}^{K_1}
      		\lrbrk{ \frac{u_{1,k}-u_{1,j}+i}{u_{1,k}-u_{1,j}-i}\, \exp \bigbrk{2i\theta(u_{1,k},u_{1,j})} }
     		\prod_{j=1}^{K_2} \frac{u_{1,k}-u_{2,j}-\ihalf}{u_{1,k}-u_{2,j}+\ihalf}\nln
\times\exp \bigbrk{2i\phi(u_{1,k})}&\lrbrk{\frac{\bar x_{1,k}^+}{\bar x_{1,k}^-}}^L
	\prod_{\textstyle\atopfrac{j=1}{j\neq k}}^{K_1}
      		\lrbrk{ \frac{\bar u_{1,k}- u_{1,j}-i}{\bar u_{1,k}- u_{1,j}+i}\, \exp \bigbrk{-2i\theta(\bar u_{1,k}, u_{1,j})} }
     		\prod_{j=1}^{K_2} \frac{\bar u_{1,k}- u_{2,j}+\ihalf}{\bar u_{1,k}-u_{2,j}-\ihalf}.
\end{align}
Here we have
\begin{align}
x_{\ell,k}^+&=x(u_{\ell,k}+\ihalf),\nln
x_{\ell,k}^-&=x(u_{\ell,k}-\ihalf),
\end{align}
and the bar represents the reflection map.
For the levels $\ell=2,\ldots,N-2$
the auxiliary Bethe equations take the standard form for $\alg{gl}(N)$
\begin{align}
\label{BetheEquationsMiddle}
1=&\prod_{j=1}^{K_{\ell-1}} \frac{u_{\ell,k}-u_{\ell-1,j}-\ihalf}{u_{\ell,k}-u_{\ell-1,j}+\ihalf}
  \prod_{\textstyle\atopfrac{j=1}{j\neq k}}^{K_\ell}\frac{u_{\ell,k}-u_{\ell,j}+i}{u_{\ell,k}-u_{\ell,j}-i}
  \prod_{j=1}^{K_{\ell+1}} \frac{u_{\ell,k}-u_{\ell+1,j}-\ihalf}{u_{\ell,k}-u_{\ell+1,j}+\ihalf}\nln
\times&\prod_{j=1}^{K_{\ell-1}} \frac{\bar u_{\ell,k}- u_{\ell-1,j}+\ihalf}{\bar u_{\ell,k}-u_{\ell-1,j}-\ihalf}
  \prod_{\textstyle\atopfrac{j=1}{j\neq k}}^{K_\ell}\frac{\bar u_{\ell,k}- u_{\ell,j}-i}{\bar u_{\ell,k}- u_{\ell,j}+i}
  \prod_{j=1}^{K_{\ell+1}} \frac{\bar u_{\ell,k}-u_{\ell+1,j}+\ihalf}{\bar u_{\ell,k}- u_{\ell+1,j}-\ihalf}
\end{align}
and for the final level $\ell=N-1$
\begin{align}
\label{BetheEquationsFinal}
1=&\prod_{j=1}^{K_{N-2}} \frac{u_{N-1,k}-u_{N-2,j}-\ihalf}{u_{N-1,k}-u_{N-2,j}+\ihalf}
  \prod_{\textstyle\atopfrac{j=1}{j\neq k}}^{K_{N-1}}\frac{u_{N-1,k}-u_{N-1,j}+i}{u_{N-1,k}-u_{N-1,j}-i}\nln
\times &\prod_{j=1}^{K_{N-2}} \frac{\bar u_{N-1,k}-u_{N-2,j}+\ihalf}{\bar u_{N-1,k}- u_{N-2,j}-\ihalf}
  \prod_{\textstyle\atopfrac{j=1}{j\neq k}}^{K_{N-1}}\frac{\bar u_{N-1,k}-u_{N-1,j}-i}{\bar u_{N-1,k}-u_{N-1,j}+i}.
\end{align}
Note that the parameters $u_{\ell,k}$ do not depend on the coupling $\lambda$.
Therefore the Bethe equations are deformed only by means of the rapidity map $x$,
the dressing phase $\theta$ and the reflection phase $\phi$.
These deformation functions are in turn characterized by
the free parameters $\alpha_\ell(\lambda)$, $\beta_{r,s}(\lambda)$
and $\delta_{2s+1}(\lambda)$, respectively, which are described below.

\section{Degrees of Freedom}

To decode the role of the different types of free coefficients
in this spin chain model it is helpful to understand
how their numbers increase with the order of $\lambda$.
Constructing the conserved charges at order $\lambda^k$ we solved the equation
\[
\sum_{l=0}^k\comm{\mathcal{H}^{(l)}}{\charge_4^{(k-l)}}=0
\]
for the undetermined parameters, i.e.\ the parameters in $\mathcal{H}^{(k)}$
and $\charge^{(k)}$. However, to determine the overall number of the
characteristic coefficients and their parity, it suffices to count the constraints
imposed by the homogeneous equation
\[\label{homeq}
\comm{\delta\mathcal{H}^{(k)}}{\charge_4^{(0)}}
+\comm{\mathcal{H}^{(0)}}{\delta\charge_4^{(k)}}=0,
\]
since they determine the dimension of the space of solutions.
To identify the parity of the parameters we simply project
out the parity even or odd part of the solutions to \eqref{homeq}
and count the remaining linearly independent structures.
These coefficients can be classified into certain
categories which are described in the following.
The numbers of free parameters in each category
are summarized in Table~\ref{coeffnumber}.

\begin{sidewaystable}
\begin{center}
\begin{tabular}{|l|c|c c|cc|cc|cc|cc|}\cline{2-10}
\multicolumn{1}{l|}{} 				&$\pm$&\multicolumn{2}{c|}{$\lambda^0$}&\multicolumn{2}{c|}{$\lambda^1$}&\multicolumn{2}{c|}{$\lambda^2$}&\multicolumn{2}{c|}{$\lambda^3$}\\ \hline
ansatz for Hamiltonian $\charge_2$		&   & $2^{2}$&$1^{1}$ & $5^{14}_{5}$&$3^{2}_{1}$ & $19^{14}_{5}$&$11^{6}_{5}$ & 								$97^{56}_{41}$&$47^{24}_{23}$ \\
ansatz for higher charge $\charge_4$		&$+$& $19^{14}_{5}$&$11^{6}_{5}$&$97^{56}_{41}$&$47^{24}_{23}$									&$601^{328}_{273}$&$239^{120}_{119}$	&$4321^{2208}_{2113}$&$1439^{720}_{719}$\\
ansatz for both 				&$=$& $21^{16}_{5}$&$12^{7}_{5}$	&$102^{60}_{42}$&$50^{26}_{24}$								&$620^{342}_{278}$&$250^{126}_{124}$	&$4418^{2264}_{2154}$&$1486^{744}_{742}$\\
constraints from commutation			&$-$& $16^{11}_{5}$&$10^{5}_{5}$&$95^{54}_{41}$&$46^{23}_{23}$ & $608^{331}_{277}$ & $242^{121}_{121}$ & $4389^{2242}_{2147}$&$1468^{734}_{734}$\\\hline
undetermined coefficients		&$=$&$5^{5}$&$2^{2}$&$7^{6}_{1}$&$4^{3}_{1}$&$12^{11}_{1}$&$8^{5}_{3}$&$29^{22}_{7}$&$18^{10}_{8}$\\
$\alpha_{\ell}$ (rapidity map)		&$-$&$0$&$-$&$2^{1}_{1}$&$-$&$2^{1}_{1}$&$-$&$4^{2}_{2}$&$-$\\
$\delta_{\ell}$ (reflection phase)		&$-$&$-$&$0$&$-$&$1^{1}$&$-$&$2^{2}$&$-$&$3^{3}$\\
$\beta_{2r,2s+1}$/$\beta_{2r+1,2s}$ (dressing factor)		&$-$&$0$&$-$&$0$&$-$&$1^{1}$&$-$&$2^{2}$&$-$\\
$\gamma_{2,s}$ (eigenvalue $\charge_2$)&$-$&$2^{2}$&$1^{1}$&$2^{2}$&$1^{1}$&$3^{3}$&$1^{1}$&$3^{3}$&$1^{1}$\\
$\gamma_{4,s}$ (eigenvalue $\charge_4$)&$-$&$3^{3}$&$1^{1}$&$3^{3}$&$1^{1}$&$4^{4}$&$1^{1}$&$4^{4}$&$1^{1}$\\\hline
$\epsilon_{k,\ell}$/$\beta_{2r,2s}$ and $\zeta_{k,\ell}$ (similarity transformations)&$=$&$0$&$0$&$0$&$1_{1}$&$2^{2}$&$4^{1}_{3}$&$16^{11}_{5}$&$13^{5}_{8}$\\
$\beta_{2r,2s}$ (bilocal similarity transformation)&$+$&$0$&$-$&$0$&$-$&$0$&$-$&$1_{1}$&$-$\\
trivial local similarity transformations		&$+$ &$1^{1}$&$0$&$2^{2}$&$1^{1}$&$2^{2}$&$1^{1}$&$3^{3}$&$1^{1}$\\
extra local similarity transformations			&$\pm$&$0$&$0$&$0$&$1_{1}$&$1_{1}$&$2_{2}$&$1_{1}$&$3_{3}$\\
all similarity transformations		&$=$&$1^{1}$&$0$&$2^{2}$&$1^{1}$&$5^{4}_{1}$&$3^{2}_{1}$&$19^{14}_{5}$&$11^{6}_{5}$\\ \hline
\end{tabular}
\end{center}
\caption{Numbers of free parameters split into bulk and boundary coefficients.
The parity of the coefficients is denoted by upper
and lower indices for even and odd parity, respectively.
The parameters $\alpha_\ell$, $\beta_{2r,2s+1}$/$\beta_{2r+1,2s}$
and $\delta_{\ell}$ characterize different spin chain systems
while the $\gamma_{r,s}$ fix linear combinations of the bare charges.
The coefficients $\epsilon_{k,\ell}$ and $\zeta_{k,\ell}$
represent local bulk and boundary similarity transformations, respectively.
The $\beta_{2r,2s}$ correspond to bilocal similarity transformations.}
\label{coeffnumber}
\end{sidewaystable}%

\paragraph{Rapidity Map.}

The \emph{rapidity map} $x(u)$ shall be defined implicitly by its inverse
\[
u(x)=x+\sum_{\ell=0}^\infty\frac{\alpha_\ell(\lambda)}{x^{\ell}}.
\]
Here the parameter functions $\alpha_\ell(\lambda)$
start at order $\mathcal{O}(\lambda^\ell)$
\[\label{alphal}
\alpha_\ell(\lambda)=\sum_{k=\ell}^\infty \lambda^k \alpha_\ell^{(k)}.
\]
Note that the leading order parameters
with even indices $\alpha_{2r}^{(2r)}$
are not free but fixed by the parameters
of the previous orders in $\lambda$
\begin{align}
\alpha_{0}^{(0)}&=0,\nln
\alpha_{2}^{(2)}&=-\alpha_{0}^{(1)} \alpha_{1}^{(1)}.
\end{align}
We believe this pattern to hold
for the higher orders as well even though we do not have
a proposal for how these parameters are fixed precisely
in terms of the lower orders.
Note that the coefficients $\alpha_{2 \ell}$
have odd parity whereas the parity of the $\alpha_{2 \ell+1}$ is even.
Hence, if the system is to conserve parity,
then one must set $\alpha_\ell=0$ for all even $\ell$.

The inverse map from the $u$-plane to the $x$-plane has the form
\[
x(u)=\frac{u}{2}+\frac{u}{2}\sqrt{1-4\sum_{\ell=0}^\infty\frac{\tilde \alpha_\ell(\lambda)}{u^{\ell+1}}}.
\]
The parameters $\tilde \alpha_\ell(\lambda)$
are fixed uniquely by the components of $\alpha_k(\lambda)$ in \eqref{alphal}.
Here $\tilde \alpha_0(\lambda)$ starts
at order $\mathcal{O}(\lambda)$ and $\tilde \alpha_{\ell\geq 1}(\lambda)$
at $\mathcal{O}(\lambda^{[l/2]+1})$.
The coefficients $\alpha_\ell(\lambda)$
govern the propagation of spin flips in the ferromagnetic vacuum.

\paragraph{Reflection Map.}

When a particle is reflected at one of the boundaries,
its momentum changes from $p$ to $\bar p(p)$.
For the first perturbative orders we find
\[\label{refmap}
\bar p(p)=
-p
+8\lambda\alpha_{0}^{(1)}\sin^2{\frac{p}{2}}
+8\lambda^2 \sin^2\frac{p}{2}
\left(\alpha_{0}^{(2)}
-4\alpha_{0}^{(1)}\alpha_{1}^{(1)} \sin^2\frac{p}{2}
-2(\alpha_{0}^{(1)})^2 \sin p
\right)
+\mathcal{O}(\lambda^3).
\]
Note that the momentum does not simply reverse its sign
but receives nontrivial corrections at higher orders in $\lambda$.
Due to the parity breaking Hamiltonian this is a necessary condition for
constructing an eigenstate of $\charge_2$.
Therefore the reflection map \eqref{refmap}
is characterized by the parity odd coefficients $\alpha_{2r}$.
The in- and outgoing momentum represent the two solutions
of the equation $E(p)=E(\bar p)$.
In our notation the reflection map $p\rightarrow \bar p$ is simply given by
\[
\bar u= -u.
\]
Note that the Bethe equations invert under $u_{\ell,k}\to \bar u_{\ell,k}$
whereas they are invariant under $u_{\ell,j}\to \bar u_{\ell,j}$
for each $j$ separately.

Recalling the definition of the elementary magnon charges
for the closed $\alg{gl}(N)$ spin chain \cite{Beisert:2005wv}
\[
q_r^\et{closed}(u)=\frac{i}{r-1}\left(\frac{1}{x(u+\frac{i}{2})^{r-1}}-\frac{1}{x(u-\frac{i}{2})^{r-1}}\right),
\]
we define the \emph{elementary magnon charges} for the open $\alg{gl}(N)$ chain by
\[
q_r(u)=\frac{1}{2}\left(q_r^\et{closed}(u)+(-1)^r q_r^\et{closed}(-u)\right).
\]
This definition is motivated by the transformation behavior under the reflection map
\begin{align}
q_{2r}(u)&=q_{2r}(-u),\nln
q_{2r+1}(u)&=-q_{2r+1}(-u).
\end{align}

\paragraph{Dressing Phase.}

The dressing phase, a common phase factor
of the bulk and boundary scattering matrices, is defined by
\[
\theta(u_1,u_2)=\sum_{r=2}^\infty\sum_{s=r+1}^\infty \beta_{r,s}(\lambda) 
\bigl(q_r(u_1) q_s(u_2)-q_s(u_1)q_r(u_2)\bigr).
\]
Here the free parameters starting at order $\mathcal{O}(\lambda^{s-1})$
are given by
\[
\beta_{r,s}(\lambda)=\sum_{k=s-1}^\infty \lambda^k\beta^{(k)}_{r,s}.
\]
The dressing phase occurs in the open Bethe equations
only in form of the difference
\[ \label{Bethedressing}
\theta(u_{1,k},u_{1,j})-\theta(\bar u_{1,k},u_{1,j}).
\]
Note that this combination inverts its sign
under $u_{1,k}\to \bar u_{1,k}$ whereas it must remain
invariant for $u_{1,j}\to \bar u_{1,j}$.
The products of two odd elementary charges $q_{2r+1} q_{2s+1}$
are not invariant under $u_{1,j}\to \bar u_{1,j}$
and are therefore not allowed
\[
\beta_{2r+1,2s+1}=0.
\]
In fact one can convince oneself that these terms
are incompatible with the boundary Yang--Baxter equation.
Furthermore, products of even elementary charges $q_{2r} q_{2s}$
drop out in the combination \eqref{Bethedressing},
i.e.\ the parameters $\beta_{2r,2s}$
with even index pairs do not appear in the Bethe equations.
This fact is related to the appearance of
bilocal similarity transformations which are described below.

\paragraph{Reflection Phase.}

The reflection phase as it appears in the above Bethe equations is defined by
\[
\phi(u)= \sum_{s=1}^{\infty} \delta_{2s+1}(\lambda) q_{2s+1}(u)-\theta(\bar u,u).
\]
The related intrinsic parameters $\delta_{2s+1}(\lambda)$
starting at $\mathcal{O}(\lambda^s)$ are then given by
\[
\delta_{2s+1}(\lambda)=\sum_{k=s}^\infty \lambda^k \delta_{2s+1}^{(k)}.
\]
These coefficients govern the scattering of particles at the boundaries.
Only the difference $2 \phi$
of the right and left boundary phase $\phi\indups{R}$ and $\phi\indups{L}$
appears as a physical parameter in the Bethe equations.
Note that the dressing part of the reflection phase
can be regarded as a $j=k$ contribution to the second line
of the $\ell=1$ Bethe equations.
Due to the transformation rules for the elementary magnon charges we find
\[
\phi(\bar u)=-\phi(u).
\]

\paragraph{Eigenvalues of the Charges.}

If we define the \emph{bare charges} by
\[
\bar Q_{2s}=\sum_{k=1}^{K_1} q_{2s}(u_{1,k}),
\]
the eigenvalues of the spin chain charges are determined by the formula
\[\label{EnergyFormula}
Q_{2r}=\gamma_{2r,-2}(\lambda) B+\gamma_{2r,0}(\lambda) L+\sum_{s=1}^\infty \gamma_{2r,2s}(\lambda)\bar Q_{2s}.
\]
The functions $\gamma_{2r,2s}(\lambda)$ are given by
\[
\gamma_{2r,2s}(\lambda)=\sum_{k=2\max(s-r,0)}^\infty \lambda^k \gamma_{2r,2s}^{(k)}.
\]
They correspond to the linear coefficients in the mixing
of the bare charges $\bar\charge_{2s}$ forming the
spin chain charges $\charge_{2r}$.
Since all bare charges commute with each other,
the coefficients $\gamma_{2r,2s}(\lambda)$ are not
shared between the spin chain charges.
All other types of coefficients are shared between the charges.

Note that in contrast to the closed spin chain parity conservation
does not impose further restrictions on the parameters $\gamma_{2r,2s}$
since only the even elementary magnon charges $q_{2s}$
contribute to the spin chain charges.
The second bare charge $\bar{\charge}_2$
is printed up to second order at the end of this paper.

\paragraph{Local Similarity Transformations.}

The coefficients $\epsilon_{k,l}$ and $\zeta_{k,l}$
do not appear in the Bethe equations and therefore
they do not influence the spectrum.
They correspond to perturbative bulk
or boundary similarity transformations of all operators
\[
\bar{\charge}_r=\mathcal{T} \tilde{\charge}_r\mathcal{T}^{-1}, \hspace{10mm} 
\mathcal{T}=1+\sum_{k=1}^\infty \lambda^k \mathcal{T}^{(k)},
\]
where $\mathcal{T}^{(k)}$ is an arbitrary bulk or
boundary interaction of range $k+1$
parametrized by $\epsilon_{k,l}$ or $\zeta_{k,l}$, respectively.
Contributions to $\mathcal{T}$ which are linear combinations
of the commuting charges do not alter the charges.
Thus for counting purposes one has to remove these
\emph{trivial similarity transformations}.

As indicated above, only the even charges $\charge_{2r}$
are conserved quantities for the open spin chain.
The charges with odd indices $\charge_{2r+1}$
can be constructed by requiring that they commute
with the Hamiltonian up to boundary terms.
Hence, structures corresponding to these odd charges do correspond
to \emph{extra similarity transformations} which have to be treated
as trivial similarity transformations
only in the bulk and there have to be removed for the counting.
For the boundary instead, the odd charges do provide nontrivial
similarity transformations and have to be added for the counting.

As an example we can commute the essential structures
corresponding to the zero-order charges $\charge_2$ and $\charge_3$
with each other
\begin{eqnarray}
 \charge_2\earel{\sim} \Perm{2,1},\nonumber\\
\charge_3\earel{\sim} \Perm{3,1,2}-\Perm{2,3,1},
\end{eqnarray}
to get
\[
 [\charge_2,\charge_3]\sim \Perm{1,3,2}-\Perm{2,1,3}.
\]
The resulting boundary structure is not of interaction range 4
as one might expect but of range 3.
As a consequence $\charge_3$ already appears
as an extra similarity transformation at order $\lambda$.
Similarly the commutator of $\charge_2$ and $\charge_5$
already appears at order $\lambda^2$ instead of $\lambda^4$
since it is of interaction range 4 instead of the expected range 6.

\paragraph{Bilocal Similarity Transformations.}

In order to match up the total numbers of parameters it is
important to consider a curious class of bilocal similarity
transformations which exists only for open chains.
Bilocal operators can be built from two local interactions
acting at different positions of the spin chain
\begin{align}
&[a_1,\dots,a_n|b_1,\dots,b_m]\ket{c_1,\dots,c_L}= \nln
&\sum_{k=1}^{L-n-m+1}\sum_{l=k+n}^{L-m+1}\ket{c_1,..,c_k,c_{k+a_1},..,c_{k+a_n},c_{k+n+1},..,c_l,c_{l+b_1},..,c_{l+b_m},c_{l+m+1},..,c_L}.
\end{align}
In a natural notation we denote by $[\charge_{2r}|\charge_{2s}]$
the bilocal composition of the two local charges $\charge_{2r}$
and $\charge_{2s}$ where $\charge_{2r}$ acts towards the
left of $\charge_{2s}$.
If we commute this particular bilocal operator
with another local even charge $\charge_{2t}$ we get a local operator.
This is due to the fact that contributions
to the commutator vanish as long as both parts of the bilocal operator
are well separated.
Only if the distance between both parts in the sum above becomes less
than the interaction range of the local charge,
the resulting interactions contribute to the commutator.
Therefore these special
\emph{bilocal similarity transformations}
give rise to additional local terms in the commuting charges.
The dressing parameters $\beta_{2r,2s}$ with even index pairs
do not appear in the Bethe equations and hence do not influence the spectrum.

\section{Tests of the Bethe Equations}

In the preceding sections we have constructed a tentatively integrable 
open spin chain Hamiltonian  and asymptotic Bethe equations to diagonalize it. 
However, we have no rigorous proof for the integrability of the system
and thus we cannot be sure that the Bethe equations are indeed correct. 

In order to test the Bethe equations \eqref{BetheEquations} we have computed
some explicit solutions for a small number of excitations on short open chains
up to second order in $\lambda$. 
Plugging these Bethe roots into \eqref{EnergyFormula} 
we have obtained the energy of the corresponding states.
On the other hand we have diagonalized the combinatorial Hamiltonian 
in Table~\ref{tab:Q2} on a suitable basis of states 
and found the spectrum to be in perfect agreement 
with the results from the Bethe equations.

\begin{table}
\<
\young{\cr}^{\phantom{1}}\eq\young{\cr}
\nln
\young{\cr}^{2}\eq
\young{\cr\cr}
+\young{&\cr}
\nln
\young{\cr}^{3}\eq
\young{\cr\cr\cr}
+2\young{&\cr\cr}
+\young{&&\cr}
\nln
\young{\cr}^{4}\eq
\young{\cr\cr\cr\cr}
+3\young{&\cr\cr\cr}
+3\young{&&\cr\cr}
+2\young{&\cr&\cr}
+\young{&&&\cr}
\nln
\young{\cr}^{5}\eq
\young{\cr\cr\cr\cr\cr}
+4\young{&\cr\cr\cr\cr}
+6\young{&&\cr\cr\cr}
+5\young{&\cr&\cr\cr}
+4\young{&&&\cr\cr}
+5\young{&&\cr&\cr}
+\young{&&&&\cr}
\nln
\young{\cr}^{6}\eq
\young{\cr\cr\cr\cr\cr\cr}
+5\young{&\cr\cr\cr\cr\cr}
+10\young{&&\cr\cr\cr\cr}
+9\young{&\cr&\cr\cr\cr}
+10\young{&&&\cr\cr\cr}
+16\young{&&\cr&\cr\cr}
+5\young{&\cr&\cr&\cr}
\nl
+5\young{&&&&\cr\cr}
+9\young{&&&\cr&\cr}
+5\young{&&\cr&&\cr}
+\young{&&&&&\cr}
\nn
\>

\caption{Decomposition of tensor products of $k$
fundamental representations of $\alg{gl}(N)$
into irreducibles.}
\label{tab:TensorYoung}
\end{table}

To be more precise, let us consider a state of length $L=4$ 
with $K_1=2$ and $K_2=1$ Bethe root excitations as an example.
It corresponds to the following Young tableau
\[
\young{&\cr\cr\cr}
\]
from the tensor product decomposition of $\young{\cr}^{4}$,
cf.\ Table~\ref{tab:TensorYoung}.
The level two Bethe root $u_{2,1}$ can be eliminated by solving
the level two Bethe equation \eqref{BetheEquationsMiddle}. 
We can then solve the Bethe equations \eqref{BetheEquations}
for the Bethe roots $u_{1,1}$ and $u_{1,2}$ 
to find three admissible solutions. 
This is done order by order in $\lambda$.
Substituting the Bethe roots into the energy formula \eqref{EnergyFormula} 
finally yields the energy of the corresponding state.

On the other hand we can apply the Hamiltonian $\charge_2$
to a basis of states characterized by the magnon numbers and
flavors we considered when solving the Bethe equations.
For our example one can take this basis to be given by the states
resulting from permutations of the entries in $\ket{1,1,2,3}$.
Note that this yields twelve eigenvalues of $\charge_2$ as opposed
to three values we want to compare to. Nine of them, however,
represent energies of the following Young tableaux
in the decomposition of $\young{\cr}^{4}$, cf.\ Table~\ref{tab:TensorYoung},
\[
\young{&&\cr\cr},\quad 
\young{&\cr&\cr},\quad 
\young{&&&\cr}.
\]

We have compared the energies calculated in these two different ways
for all states corresponding to the following Young tableaux in
the tensor product decomposition illustrated in Table~\ref{tab:TensorYoung}:
\[\begin{array}[b]{c}
\young{\cr\cr\cr},\quad
2\young{&\cr\cr},\quad
3\young{&\cr\cr\cr},\quad
3\young{&&\cr\cr},\quad
2\young{&\cr&\cr},\quad
4\young{&&&\cr\cr},
\\[0.5cm]
5\young{&&\cr&\cr},\quad
6\young{&&\cr\cr\cr},\quad
5\young{&&&&\cr\cr},\quad
9\young{&&&\cr&\cr}.
\end{array}
\]
Our tests confirm that the Bethe equations describe the spectrum of $\charge_2$
correctly giving us confidence that the derived spin chain Hamiltonian
is indeed integrable.

\section{Finite Length Contributions}
\label{finiteL}

Let us discuss briefly the issue of \emph{spanning} terms \eqref{flform}.
These span the whole length of a finite open chain
in analogy to the wrapping terms for closed chains,
cf.\ \cite{Beisert:2004ry,Sieg:2005kd}.
They can be represented in the same basis of local operators
and thus a complete enumeration of such terms is easily feasible
by not identifying any of the interaction terms \eqref{eq:terms}.
We have repeated the analysis of Section~\ref{sec:constr}
including the spanning terms. We have found no restriction on
the bulk or boundary terms,
but there are additional spanning terms.
We will not specify these terms explicitly, but merely
present the number of such terms in Table~\ref{tab:flterm}.

Let us compare the numbers
to the set of irreducible representations of $\alg{gl}(N)$
for a chain of length $L$ in Table~\ref{tab:TensorYoung}
(we assume that $N\geq L$).
We write the decomposition formally as
\[
\rep{N}^L=\sum\nolimits_k n_{L,k} \rep{R}_{L,k},
\]
where $n_{L,k}$ are the multiplicities of the
irreducible representations $\rep{R}_{L,k}$.
Now it turns out that the number
of irreducible components of the tensor product
\[
\sum\nolimits_{k}
n_{L,k}=
1,2,4,10,26,76,\ldots
\quad\mbox{for}\quad
L=1,2,3,4,5,6,\ldots
\]
appears to match precisely with the number of
independent coefficients
affecting the eigenvalues of $\charge_{2,4}$ in Table~\ref{tab:flterm}.
This means that starting from
a certain perturbative order
the eigenvalue spectrum is \emph{freely adjustable}.
The \emph{spanning order} for
$\charge_{2r}$ at length $L$ is given by
$\lambda^{L+2-2r}$.
In other words,
our version of perturbative integrability
completely loses its constraining power at finite length.
Equivalent claims have been made in Section 6.5.4.\ of
\cite{Beisert:2004ry} for wrapping interactions
in closed chains.

Before we comment on the implications of this result, let us discuss the
remaining finite length parameters labeled by S.
These appear first at order $\lambda^L$ where the spectrum
of \emph{all} charges $\charge_{2r}$ becomes fully adjustable.
In that case, also similarity transformations
of the space of chains of length $L$ are permissible.
The number of such similarity transformations is given by the formula
\[
\sum\nolimits_{k}
n_{L,k}(n_{L,k}-1)=
L!-\sum\nolimits_{k}n_{L,k}
=
0,0,2,14,94,\ldots
\quad\mbox{for}\quad
L=1,2,3,4,5,\ldots\,.
\]
It is based on the fact that similarity transformations
can only act on identical irreducible representations.
For $n$ identical irreps, there are $n^2$ similarity
transformations, but $n$ of them act trivially on a
given set of commuting matrices $\charge_{2r}$.
Furthermore the number
of similarity transformations
is equal to the number of
spanning interactions of length $L$ which
equals $L!$.

\begin{table}\centering
\begin{tabular}{|c|rr|rr|rr|rrr|rrr|}\hline
   &
\multicolumn{2}{c|}{$\lambda^0$}&
\multicolumn{2}{c|}{$\lambda^1$}&
\multicolumn{2}{c|}{$\lambda^2$}&
\multicolumn{3}{c|}{$\lambda^3$}&
\multicolumn{3}{c|}{$\lambda^4$ predicted}\\
$L$&$\charge_2$&$\charge_4$
   &$\charge_2$&$\charge_4$
   &$\charge_2$&$\charge_4$
   &$\charge_2$&$\charge_4$&S
   &$\charge_2$&$\charge_4$&S\\\hline
1&  &1&   1&1&   1& 1&   1& 1& &  1& 1&  \\
2&  &2&    &2&   2& 2&   2& 2& &  2& 2&  \\
3&  & &    &4&    & 4&   4& 4&2&  4& 4& 2\\
4&  & &    & &    &10&    &10& & 10&10&14\\
5&  & &    & &    &  &    &26&2&   &26& 0\\
6&  & &    & &    &  &    &  & &   &76& ?\\
\hline
\end{tabular}
\caption{Number of spanning terms at given order and length.
Terms listed under $\charge_{2,4}$
affect the eigenvalues of $\charge_{2,4}$, respectively.
Terms listed under S correspond to similarity transformations.}
\label{tab:flterm}
\end{table}

Exceptionally we find a non-zero entry in Table~\ref{tab:flterm}
for similarity transformations at $L=5$ and $\lambda^3$.
It is explained by a degeneracy of two eigenvalues of $\charge_2$
at leading order for identical irreps at $L=5$
(the one with multiplicity 6 in Table~\ref{tab:TensorYoung}).
In addition to adjusting these two eigenvalues
one can also adjust the eigenvectors
with $2(2-1)=2$ similarity transformations.

How can we understand the complete arbitrariness
of finite length contributions?
At spanning order for open chains
(and analogously at wrapping order for closed chains)
the interactions act on the whole chain. One thus loses
the desired \emph{locality} property of $\charge_{r}$
and \emph{all linear operators} become admissible contributions.
We may thus consider the charges $\charge_{r}$
as free general matrices. In the generic case,
two $n\times n$ matrices commute if their eigenvectors coincide.
Therefore one should expect $2n$ degrees of freedom for the eigenvalues
of the two matrices and $n(n-1)$ degrees of freedom for the common $n$ eigenvectors
(up to rescaling).

What implications does this observation have on integrability
of (perturbative) long-range chains? There are two possibilities
which crucially depend on the precise definition of integrability
(which is currently unclear):
In the one case, integrability permits all the
finite length terms that we have found.
This implies that the operator $\charge_2$
is almost completely arbitrary at sufficiently high orders
of $\lambda$ or at finite $\lambda$.
Clearly there cannot be a method to determine the spectrum
which is remotely as efficient as the Bethe ansatz.
Therefore this definition of integrability cannot be
useful in practice.

There may be various stronger definitions of integrability
which would constrain some or all of the
spanning (wrapping) terms.
In addition they might even constrain some
of the bulk (or boundary) terms.
The last option would be particularly desirable
for AdS/CFT integrability because it would provide
us with further constraints on the finite size spectrum.
Consider for example the Haldane--Shastry and Inozemtsev
integrable long-range chains
\cite{Haldane:1988gg,SriramShastry:1988gh,Inozemtsev:1989yq,Inozemtsev:2002vb}.
Their finite size spectrum is uniquely defined by the Hamiltonian.
More conveniently it can be described exactly
by equations similar to Bethe equations
(for Haldane--Shastry the equations are purely combinatorial
while for Inozemtsev the equations are currently known only in special cases).
It appears unlikely that the equations can be deformed in such a way
that only the spectrum at spanning or wrapping order is deformed.
Hopefully a suitable integrability constraint can be found to
constrain the spectrum of $\mathcal{N}=4$ SYM at finite size.

\section{Summary}

In this paper we have generalized the analysis of long-range integrable $\alg{gl}(N)$
spin chains started in \cite{Beisert:2005wv} to the case of open boundary conditions.
For these open chains only the even integrable charges $\charge_{2r}$ are conserved
and we have explicitly constructed the first two of them up to order $\lambda^3$.
We have performed the nested asymptotic Bethe ansatz for the open spin chain
and found several new features characterizing its moduli space:
\begin{itemize}
 \item
a new type of deformation parameter $\delta_{2s+1}$
starting at $\mathcal{O}(\lambda^s)$ which represents 
a phase for reflections of particles at the boundaries of the chain,

 \item
a reflection map $\bar p(p)$ corresponding
to a nontrivial change of particle momentum at the boundaries
due to a parity breaking Hamiltonian,

 \item
boundary and bilocal similarity transformations having
their origin in the open boundary conditions.

\end{itemize}
On a set of states we have demonstrated that the Bethe equations
yield the correct energy spectrum which indicated that the 
proposed Hamiltonian is indeed integrable.
In addition we have briefly discussed the role of \emph{spanning}
terms in the context of open chains as the counterpart
of \emph{wrapping} interactions for closed spin chains.
We found that these finite size contributions
do not impose any restrictions on the bulk or boundary interactions
but merely provide additional degrees of freedom not influencing
the integrability of the system.

\subsection*{Acknowledgments}

We are grateful to Abhishek Agarwal for his collaboration at earlier stages of this project.

\bibliographystyle{nb}
\bibliography{OpenChain}

\vspace{\fill}

\begin{table}[ht]
\begin{align}
\bar {\cal Q}_2(\lambda)=&\mathrel{}(\Perm{1, 2} - \Perm{2, 1})\nn\\[7pt]
				&+i \alpha_0(\lambda)\, ( \Perm{2, 3, 1} - \Perm{3, 1, 2})\nln
				&+\alpha_1(\lambda)\, (\Perm{1} - 4 \Perm{1, 2} + 2 \Perm{2, 1} + \Perm{1, 3, 2} + \Perm{2, 1, 3} - \Perm{3, 2, 1})\nln
				&+\half\delta_3(\lambda)\, (2 \Perm{1} - 2 \Perm{1, 2} - 2 \Perm{2, 1} + \Perm{1, 3, 2} +\Perm{2, 1, 3})\nln
				&+ \zeta^-_{1,1}(\lambda)\, (-\Perm{1, 3, 2} + \Perm{2, 1, 3})\nn\\[7pt]
				&+2\alpha_0(\lambda)^2\,  ( \Perm{1} - 2 \Perm{1, 2} + \Perm{1, 3, 2} + \Perm{2, 1, 3} - \Perm{3, 2, 1}\nln
				& 	\qquad + \Perm{2, 3, 4, 1} - \Perm{2, 4, 1, 3} - \Perm{3, 1, 4, 2} + \Perm{4, 1, 2, 3})\nln
				&+\alpha_1(\lambda)^2\, (-12 \Perm{1} + 32 \Perm{1, 2} - 7 \Perm{2, 1} - 11 \Perm{1, 3, 2} - 11 \Perm{2, 1, 3} + 8 \Perm{3, 2, 1} \nln
				&	\qquad	+ \Perm{1, 4, 3, 2} - \Perm{2, 3, 4, 1} + \Perm{2, 4, 1, 3} + \Perm{3, 1, 4, 2}\nln
				&	\qquad + \Perm{3, 2, 1, 4} - \Perm{4, 1, 2, 3} - \Perm{4, 2, 3, 1})\nln
				&+\ihalf \alpha_0(\lambda) \alpha_1(\lambda)\, (-18 \Perm{2, 3, 1} + 18 \Perm{3, 1, 2} + 3  \Perm{2, 4, 3, 1} + 3  \Perm{3, 2, 4, 1}\nln
				&	\qquad - 3  \Perm{4, 1, 3, 2} - 3  \Perm{4, 2, 1, 3})\nln
				&+\sfrac{1}{8} \delta_3(\lambda)^2\, (-2\Perm{2, 1}  - 2\Perm{3, 2, 1}+ 3  \Perm{1, 3, 2} + 3 \Perm{2, 1, 3} - 2\Perm{1, 2, 4, 3}\nln
				&	\qquad  - 2\Perm{2, 1, 3, 4} + \Perm{1, 4, 3, 2}+  \Perm{3, 2, 1, 4})\nln
				&+\sfrac{1}{4}\delta_5(\lambda)\, (-4 \Perm{1} +4 \Perm{1, 2} + 2\Perm{2, 1}  -2\Perm{3, 2, 1}+  \Perm{1, 3, 2} +  \Perm{2, 1, 3}\nln
				&	\qquad - 2\Perm{1, 2, 4, 3}- 2\Perm{2, 1, 3, 4} + \Perm{1, 4, 3, 2} + \Perm{3, 2, 1, 4})\nln
				&+ \alpha_1(\lambda) \delta_3(\lambda)\, (-10 \Perm{1} + 10 \Perm{1, 2} + 10 \Perm{2, 1} - 4 \Perm{1, 3, 2} - 4 \Perm{2, 1, 3} - 2 \Perm{3, 2, 1}\nln
				&	\qquad- \Perm{1, 2, 4, 3} + \Perm{1, 4, 3, 2} - \Perm{2, 1, 3, 4} + \Perm{3, 2, 1, 4})\nln
				&+\half \beta_{23}(\lambda)\, (-4 \Perm{1} + 8 \Perm{2, 1} - 2\Perm{2, 3, 1} - 2\Perm{3, 1, 2} - 2\Perm{2, 1, 4, 3} - 2\Perm{2, 3, 4, 1}\nln
				&	\qquad+ 2\Perm{2, 4, 1, 3}  + 2\Perm{3, 1, 4, 2} - 2\Perm{3, 4, 1, 2} - 2\Perm{4, 1, 2, 3}\nln
				&	\qquad +  \Perm{2, 4, 3, 1} +  \Perm{3, 2, 4, 1}+  \Perm{4, 1, 3, 2} + \Perm{4, 2, 1, 3})\nln
				&+i \epsilon^+_{2,1}(\lambda)\, ( \Perm{2, 4, 1, 3} - \Perm{3, 1, 4, 2})\nln
				&+i \epsilon^+_{2,2}(\lambda)\, ( \Perm{2, 4, 3, 1} - \Perm{3, 2, 4, 1} - \Perm{4, 1, 3, 2} + \Perm{4, 2, 1, 3})\nln
				&+\half\zeta^-_{1,1}(\lambda)^2\, (-2\Perm{2, 1} + 3 \Perm{1, 3, 2} + 3 \Perm{2, 1, 3} - 2\Perm{3, 2, 1} - 2\Perm{1, 2, 4, 3}  - 2\Perm{2, 1, 3, 4}\nln
				&	\qquad +  \Perm{1, 4, 3, 2}+  \Perm{3, 2, 1, 4})\nln
				&+\zeta^-_{2,1}(\lambda)\, (2 \Perm{1, 2, 4, 3} - 2 \Perm{2, 1, 3, 4}- \Perm{1, 4, 3, 2}  + \Perm{3, 2, 1, 4})\nln
				&+i \zeta^-_{2,2}(\lambda)\, (2  \Perm{2, 3, 1} - 2  \Perm{3, 1, 2} - \Perm{1, 3, 4, 2} + \Perm{1, 4, 2, 3} - \Perm{2, 3, 1, 4} + \Perm{3, 1, 2, 4})\nln
				&+i \zeta^+_{2,1}(\lambda)\, (\Perm{1, 3, 4, 2} - \Perm{1, 4, 2, 3} - \Perm{2, 3, 1, 4} + \Perm{3, 1, 2, 4})\nln
				&+\alpha_1(\lambda) \zeta^-_{1,1}(\lambda)\, (-2 \Perm{1, 2, 4, 3} + 2 \Perm{2, 1, 3, 4})\nn\\[7pt]
				&+{\cal O}(\lambda^3)\nn
\end{align}
\caption{Normalized Hamiltonian printed up to second order.}
\label{tab:Q2}
\end{table}


\end{document}